\documentclass[prd,twocolumn,showpacs,aps,superscriptaddress]{revtex4-1}

\usepackage{graphicx}
\usepackage{latexsym}
\usepackage{amssymb}
\usepackage{amsmath}
\usepackage[english]{babel}

\begin{document}

\author{V.S. Berezinsky}\thanks{e-mail: berezinsky@lngs.infn.it}
\affiliation{INFN, Laboratori Nazionali del Gran Sasso, Center for Astroparticle Physics at LNGS (CFA), \\ I-67010 Assergi (AQ), Italy}
\title{Formation and internal structure of superdense dark matter clumps and
ultracompact minihaloes}
\author{V.I. Dokuchaev}\thanks{e-mail: dokuchaev@lngs.infn.it}
\author{Yu.N. Eroshenko}\thanks{e-mail: eroshenko@inr.ac.ru}
\affiliation{Institute for Nuclear Research of the Russian Academy of
Sciences \\ 60th October Anniversary Prospect 7a, 117312 Moscow, Russia }

\date{\today}

\begin{abstract}
We discuss the formation mechanisms and structure of the superdense
dark matter clumps (SDMC) and ultracompact minihaloes (UCMH),
outlining the differences between these types of DM objects. We
define as SDMC the gravitationally bounded DM objects
which have come into virial equilibrium at the radiation-dominated
(RD) stage of the universe evolution. Such objects can be formed from the
isocurvature (entropy) density perturbations or from the peaks in the
spectrum of curvature (adiabatic)  perturbation. The axion
miniclusters (Kolb and Tkachev 1994) are the example of the former
model. The system of central compact mass (e.~g. in the form of SDMC
or primordial black hole (PBH)) with the outer DM envelope formed in
the process of secondary accretion we refer to as UCMH. Therefore, the
SDMC can serve as the seed for the UCMH in some scenarios. Recently,
the SDMC and UCMH were considered in the many works, 
and we try to systematize them here. We
consider also the effect of asphericity of the initial density
perturbation in the gravitational evolution, which decreases the SDMC
amount and, as the result, suppresses the gamma-ray signal from DM
annihilation.
\end{abstract}

\maketitle

\section{Introduction}

Although the existence of DM is reliably established, the nature of
the DM particles is still unknown. There are several well motivated DM
models, which allow the direct and indirect experimental verification
in the near future. The most bright indirect signature is the
particles annihilation, and this process could be boosted inside
the dense DM clumps.

In the standard scenario the DM clumps are formed from the CMB-normalized
power-law spectrum of perturbations.  The clumps participate in the
hierarchical clustering, and finally have near power-law mass
spectrum, obtained both in the numerical simulations
\cite{DieMooSta05} and analytically \cite{BerDokEro08}. In the case of
neutralino DM the minimum mass of the clumps is $M_{\rm
min}\sim10^{-6}M_\odot$     (see e.~g. \cite{GreHofSch05}, \cite{Bri09}).   These lightest clumps are formed at the red-shift
$z\sim50$. The situation could be very different for other DM particle
candidates, e.~g. in the case of the superheavy DM particles
\cite{pI}, and/or for the non-standard spectrum of perturbations. DM
clumps can be formed very early in the beginning of the matter-dominated or
even at the RD-stage from the large density perturbations of different
nature: peaks in the spectrum of perturbations generated during
inflation or at phase transitions in the early universe, or with the
PBHs or cosmic string loops as the seeds. In this paper we consider in
some details the two somehow related types of DM clumps: SDMC and
UCMH.

In \cite{KolTka94}  the formation of SDMCs at the RD-stage evolution
of the universe from the entropy perturbations (disturbances of matter
content with constant curvature)  was considered. The \cite{DokEro02}
generalized this formalism for the adiabatic perturbations, and it was
found that the PBHs and SDMCs can be connected due to their common
origin from the same spectrum of density perturbations. SDMC forms if
the perturbation is not sufficient for formation of PBH. Such
SDMCs can be figuratively called as ``failed black holes'', because in
the case of higher perturbation the PBH would be formed instead of
SDMC. A similar idea was discussed in
\cite{ScoSiv09},   \cite{RicGou09},   \cite{JosGree10}, \cite{Yanetal11},
\cite{Yanetal12-1}, \cite{Yanetal12-2}, and in more detail -- in
\cite{BriScoAkr11}. It is important that the observational cosmic
gamma-rays constraints can limit the DM annihilation and the spectrum
of initial perturbations \cite{JosGree10},
\cite{BriScoAkr11},   \cite{Shaetal12}.   However, to establish reliably  
such  restrictions, it is necessary to determine the structure of SDMC,
especially in its central region.

Later, at the matter-dominated stage, the DM concentrates around the SDMCs in
the process of secondary accretion \cite{Ber85} forming the halo with
density profile $\propto r^{-9/4}$ \cite{Ber85}, \cite{KolTka94}. The
similar halos can grow around PBHs \cite{DokEro01}, \cite{DokEro03},
\cite{MacOstRic07},   \cite{LacBea10}, \cite{Zha11}   \cite{SaiShi11}.     The DM clumps, formed on the dust-like stage,
probably have the density profiles of the Gurevich-Zybin type
$r^{-1.8}$ \cite{GurZyb8895-1}, \cite{GurZyb8895-2}, \cite{GurZyb8895-3}, if 
these objects are isolated. Gurevich-Zybin theory explains the formation of 
singular density profile from the initially smooth perturbation. The
Navarro-Frenk-White (NFW) profile is applicable, if the objects   were not 
isolated during their formation but   experienced hierarchical clustering. 
Such a transition from NFW to a steeper inner profile near the minimum 
mass scale (objects at this scale are formed almost as isolated) 
was observed in the work \cite{AndDie13}.   Gurevich-Zybin profile should 
give smaller annihilation signal compared with the calculations 
\cite{BriScoAkr11}, where the $r^{-9/4}$ profile was used.  
The simulations of \cite{Vogetal09} indicate in favor of the 
$r^{-9/4}$ Fillmore \& Goldreich / Bertschinger secondary infall
profile 
even for isolated halos forming without a compact central object. Therefore more work is needed to clarify this topic.  

If the UCMH formed around a superdense SDMC, which virialised at the
RD stage, the central density is determined by the SDMC's
structure. First, the  calculation of the average density of SDMC
requires the formalism of \cite{KolTka94}, \cite{DokEro02},
\cite{pI}. Second, the Gurevich-Zybin   or steeper   profile probably 
is formed in the
SDMCs because of their isolation. Third, the question about the radius
of the central core has not yet definite solution. Really, the central
density of halos and subhalos is one of the unresolved problem in the
observations and in the theory of the hierarchical DM structure
formation.

Using the conventional models of WIMPs annihilation,
in the papers \cite{JosGree10}, \cite{Yanetal12-1}, \cite{Yanetal12-2}, \cite{BriScoAkr11} it was concluded from the absence of the observed annihilation signal, that the SDMCs 
can constitute only $\ll1$ fraction of DM. A similar result was 
obtained in \cite{Yanetal11} from
the influence of the early annihilation on recombination and taking
into account the known limitations to this effect from the CMB
observations.   The influence of UCMHs on the intergalactic medium and 
reionisation of the universe was studied in details by \cite{Zha11}. 
The restrictions on the UCMHs in the decaying dark mater model were 
considered in \cite{YanYanZon13}. Some restrictions were also 
obtained from the upper limits on the neutrino signals from 
annihilation \cite{YanYanZon13-2}.   The lack of the observed 
gamma-ray point sources demonstrates that in the epoch 
of $e^+e^-$ annihilation the value of perturbations at the
horizon crossing was less than $10^{-3}$ \cite{ScoSiv09}. The SDMC, if
they exists, were not destroyed in the tidal interactions during the
formation of structures at large scales. One may propose the
non-sphericity of the density perturbations as the effects that could
suppress the formation and diminish the amount of the SDMC. In this
paper we will show that non-sphericity plays the important role,
and the SDMC number can be suppressed by several orders of magnitudde.
In the papers \cite{ScoSiv09}, \cite{JosGree10}, \cite{Yanetal11},
\cite{SaiShi11}, \cite{BriScoAkr11}, \cite{Yanetal12-1},
\cite{Yanetal12-2} the maximum density in the center of the UCMH is
found from the effect of the particles annihilation. It was assumed
that since the formation moment of UCMH the more and more 
particles have time to annihilate, the region of the annihilation expands, and therefore the
radius of the core increases. This approach goes back to \cite{Ulletal02}.
In the opposite approach \cite{BerGurZyb92}, \cite{BerBotMig96},
proposed earlier, it was taken into account that in addition to the
annihilation the continuous stream of particles into the center of the
UCMH exists, so that the core radius does not change significantly
over time.

In addition to the annihilation there are other restrictions on the central
density and radius of the core. In \cite{BriScoAkr11} the core radius $R_c$ was estimated by
considering the transverse velocities which resulted in
$R_c/R\sim3\times10^{-7}$, where $R$ is the virial
radius. Restrictions on the phase space density come from the
Liouville theorem, from the tidal forces \cite{BerDokEro03},
from the presence of  substructures and the other 
sources of entropy generation
\cite{DorLukMik12}. It is likely that the   clumps    have the core radius
with $10^{-3}-10^{-2}$ fraction of $R$ as it follows from the
numerical simulations of \cite{DieMooSta05} (Fig.~2).

The important question is the final mass of the growing UCMH. 
In the papers \cite{MacOstRic07}, \cite{ScoSiv09}, \cite{Yanetal11},
\cite{SaiShi11}, \cite{Yanetal12-1}, \cite{Yanetal12-2} it was assumed
that the ceasing of the mass growth during the secondary accretion
occurred at $z\sim10-30$, when the large-scale structure began forming
actively. Previously, a rigorous criterion for the end of the growth
was found in the papers \cite{DokEro01}, \cite{DokEro03} and it also
will be discussed in this paper.


\section{Formation of superdense clumps}
\label{formsec} 

For the clumps to be formed at the RD-stage of the universe evolution,
the spectrum of perturbations must have an excess at small scale. The
ordinary CMB-normalised power-low spectrum with the observed power
index $n_s\simeq0.96$ provides too small perturbations at the RD-stage
for the clumps formation there. Really, the spectrum of the curvature
perturbations in this case is \cite{GorRub10}
\begin{equation}
{\mathcal P}_{\mathcal R}=A_{\mathcal
R}\left(\frac{k}{k_*}\right)^{n_s-1},
\label{prspectrum}
\end{equation}
where $k_*/a_0=0.002$~Mpc$^{-1}$, $A_{\mathcal
R}=(2.46\pm0.09)\times10^{-9}$, and $n_s=0.960\pm0.014$.  Therefore,
the typical amplitude of the perturbations $\Delta_{\mathcal
R}\simeq5\times10^{-5}$ is insufficient for the clumps formation at
the RD-stage.

For the further examples we consider the peak-like excess at some
scale superimposed on the ordinary power low spectrum
(\ref{prspectrum}) of adiabatic perturbations. In the case of the
sufficiently high peak, the clumps can be formed even at the RD-stage. The
constraints on the possible parameters of the peak, its position at
CDM mass scale $M$ and the peak's high $\delta_H$, are imposed by the
PBH overproduction limits.

  \subsection{Spherical collapse}
\label{sphsubsec}  

Let us outline  formalism of the SDMCs formation at the RD-stage,
which was developed in \cite{KolTka94} for entropy perturbations and
generalized for the curvature perturbations in \cite{DokEro02}. We
will start from the linear stage of evolution of the density 
perturbations.
We will use the following notation: $x=k\eta$, where $k$ is the 
co-moving wave vector of
the perturbation, which could be expressed through DM mass $M$ of the
SDMC and $\eta$ is the conformal time with $d\eta=cdt/a(t)$. 
The adiabatic DM
perturbation at $x\gg 1$ follows the law \cite{GorRub10}
\begin{equation}
\delta=-9\Phi_i\left[ \ln\left(\frac{x}{\sqrt{3}}\right)+{\bf
C}-\frac{1}{2} \right],\label{dgame}
\end{equation}
where $\Phi_i$ is the initial gravitational potential in the conformal
Newtonian frame, which is related to the perturbation of curvature
and to the perturbation of the radiation density at the horizon
crossing as $\Phi_i=-2{\mathcal R}/3\approx-0.2\delta_H$, and  ${\bf
C}\simeq0.577$ is the Euler constant.  The (\ref{dgame}) serves as the
initial condition for the evolution of DM density perturbation at the
linear stage, then $\delta\ll1$, according to the equation
\cite{GorRub10}
\begin{equation}
y(y+1)\delta''+\left(1+\frac{3}{2}y\right)\delta'-\frac{3}{2}\delta=0,\label{bigeqlin}
\end{equation}
where $y=a(\eta)/a_{\rm eq}$, and the prime denotes the derivative
over $y$. This linear stage gives the initial conditions for the
further nonlinear evolution.

Suppose that there is a positive density perturbation $\delta (\vec
r)$. The origin of coordinates $\vec r=0$ is chosen near the center of
mass of the protohalo. In the first approximation the perturbation can
be regarded as the spherically symmetric object. Denote by $M$ the DM
mass within some spherical layer. The contribution of the pressure of
the homogeneous relativistic component into the energy-momentum tensor
can be taken into account by replacing $\rho\to  \rho+3pc^2$. Then,
the evolution of a spherical layer at the sub-horizon scales $r\ll ct$
obeys the equation
\begin{equation}
\frac{d^2r}{dt^2}=-\frac{G(M_h+M)}{r^2}-\frac{8\pi G  \rho_r r}{3} +
\frac{8 \pi G\rho_{\Lambda}r}{3} \label{d2rdt1}
\end{equation}
It is taken into account the possibility of a bare mass  $M_h$, such
as the mass of the black hole at the center of the layer. One may
use the following parametrization:
\begin{equation}
r=a(y)b(y)\xi,
\label{abxi}
\end{equation}
where $\xi$ is the comoving coordinate.  Then $b$ obeys the equation
\cite{KolTka94}
\begin{equation}
y(y+1)b''+\left(1+\frac{3}{2}y\right)b'+\frac{1}{2}
\left(\frac{1+\delta_i}{b^2}-b\right)=0.\label{bigeq}
\end{equation}
The Eq.(\ref{bigeq}) was solved in \cite{KolTka94} for isocurvature
(entropic) perturbations. In this case the initial velocity
$db/dy\simeq0$, and the perturbation grows due to the initial value of
the perturbation $\delta_i\neq0$. According to \cite{KolTka94} the
central density of the clump is $140\rho_{\rm
eq}\delta_i^3(1+\delta_i)$, although our calculations gives for the
mean density the smaller value $\simeq17\rho_{\rm eq}\delta_i^4$.

\begin{figure}[t]
\begin{center}
\includegraphics[angle=0,width=0.45\textwidth]{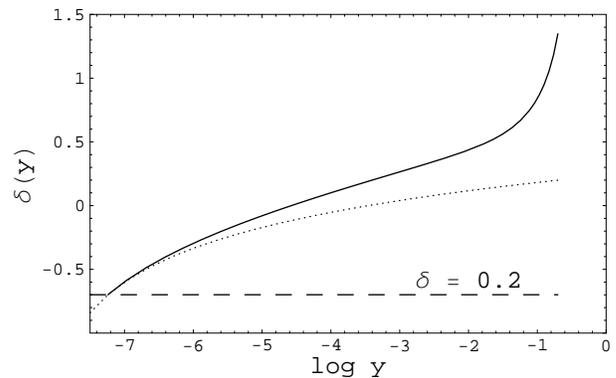}
\end{center}
\caption{Example of the SDMC evolution with $M=10^{-6}M_\odot$ and
$\delta_H=0.05$. The solid curve was obtained by the numerical
solution of (\ref{bigeq}) with the initial conditions according to
(\ref{dgame}), matched at $\delta=0.2$ (dashed curve). The dotted
curve shows the extrapolation of the linear law (\ref{dgame}) to the
large times.} \label{grex}
\end{figure}

For the adiabatic perturbations the nonzero initial quantity $db/dy$ is
specified according to the solution (\ref{dgame}). To link the
Eulerian and Lagrangian approaches, we put $b_i=(1+\delta_i)^{-1/3}$
at some  initial $y_i$. For $x\gg1$ and $y\ll1$, we have the
connection \cite{DokEro02}
\begin{equation}
x=\frac{\pi}{2^{2/3}}\left(\frac{3}{2\pi}\right)^{1/6}
\frac{yc}{M^{1/3}G^{1/2}\rho_{\mathrm{eq}}^{1/6}},
\label{xy}
\end{equation}
then
\begin{equation}
\left.b'\right|_{y_i}=-\frac{\delta_Hb_i^4} {2y_i\phi},
\end{equation}
where $\phi\approx0.817$.
The Eq. (\ref{bigeq}) can be solved numerically. The moment $y_i$ is
chosen in the region where (\ref{dgame}) and (\ref{bigeq}) are both
valid. By fixing $\delta_i=0.2$, we obtain $x_i$ and $y_i$ from the
equations (\ref{dgame}) and (\ref{xy}). The particular example of the
$\delta=b^{-3}-1$ evolution is illustrated at Fig~\ref{grex}.

The expansion of SDMC stops when $dr/dt=0$, which is equivalent to
$db/dy=-b/y$ \cite{KolTka94}. Let us use the index ``max'' for the
quantities at the the moment of the stop, then the density and radius
of the  SDMC at this time are
\begin{equation}
\rho_{\rm max}=\rho_{\mathrm{eq}}y_{\mathrm{max}}^{-3}
b_{\mathrm{max}}^{-3}, \qquad R_{\rm
max}=\left(\frac{3M}{4\pi\rho_{\mathrm{max}}} \right)^{1/3}.
\end{equation}
After virialization the final radius $R=R_{\rm max}/2$ and therefore
$\bar\rho=8\rho_{\rm max}$.   Virialization is the DM mixing inside
the forming halo till the equilibrium state. It works due to the
presence of radial osculations, large irregularities and non-spherical
movements. In the terminology of Lynden-Bell \cite{Lynden-Bell67},
these processes are called `` violent relaxation''.  The hierarchical
clustering and associated relaxation also produce the universal
density profiles, such as Navarro-Frenk-White profile \cite{Whi96}.
The new feature of the vitialization at the RD-stage is the
disintegration of the very irregular protohalo. Indeed, if the
protohalo is sufficiently elongated then the mass of the radiation
inside the enveloping sphere will exceed the mass of DM, and the
self-gravitation of the object will not hold the protohalo's parts
together. The protohalo will decay and the parts of it will fly away
one after another. To obtain the boundary of this regime of evolution,
leading to the decay, the numerical simulations of SDMC at the
RD-stage are desirable.

\begin{figure}[t]
\begin{center}
\includegraphics[angle=0,width=0.45\textwidth]{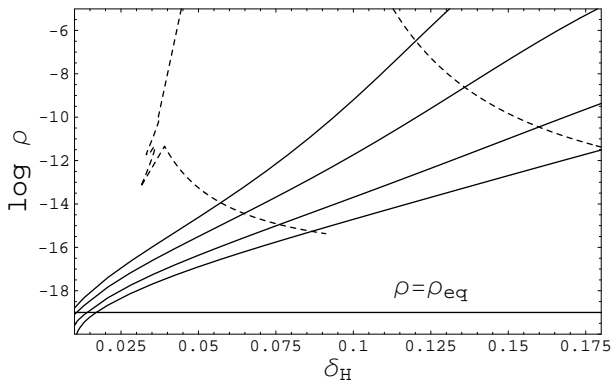}
\end{center}
\caption{Density of the SDMC in dependence on the perturbation value
$\delta_H$ for different clumps' masses $M=10^{-11}$, $10^{-6}$,
$10^{-1}$, $10^{2}M_\odot$ (from up to down). The PBH restrictions is
shown for the thresholds of the PBH's formation $\delta_{\rm th}=1/3$
and $\delta_{\rm th}=0.7$ (bottom and top dashed curves,
respectively). The local minimum on the dashed curve corresponds to
the Hawking evaporating PBHs.} \label{grpbh}
\end{figure}

The SDMCs are formed due to the beginning of the matter dominated stages
inside the local areas of the space, while on the average the universe
is still at the RD-stage. The calculated $\bar\rho$ is shown at the
Fig~\ref{grpbh}.

The SDMCs are restricted by the process of the PBH formation. The mass
of DM in some spherical volume is connected to the mass of the
radiation at the moment of the horizon crossings as \cite{DokEro02}
\begin{equation}
M_{\rm
H}\simeq2\times10^5\left(\frac{M}{0.1M_{\odot}}\right)^{2/3}M_{\odot},
\label{mhmx}
\end{equation}
therefore the PBHs restrict the SDMCs from the different mass
region. The PBHs are formed in the tail of the Gaussian perturbations
distribution, but SDMC are formed from perturbations of the rms
order. The PBHs and SDMC are connected not individually but through
their common spectrum of perturbation. As it's clear from the
Fig~\ref{grpbh}, the PBH restriction are very sensitive to the
threshold $\delta_{\rm th}$ of the PBH formation. According to
different calculations this threshold is $\delta_{\rm th}=1/3-0.7$,
where $\delta_{\rm th}=1/3$ corresponds to the simple analytic
arguments of B.J.~Carr, and the numerically revealed critical
gravitational collapse gives even $\delta_{\rm th}=0.7$
\cite{NieJed99} . There is only small room for the SDMCs in the case
$\delta_{\rm th}=1/3$.

The small-scale clumps from high peaks form earlier in comparison with
the larger objects. Therefore the small  clumps are not destructed in
the hierarchical clustering.

For illustration we consider the three particular examples, which are
presented in the Table~\ref{tab}:

1. The clumps with minimum mass $M\simeq10^{-6}M_\odot$ in the case of
   the  standard spectrum of perturbation, normalized on the 7-year
   WMAP data (power index of the primordial spectrum
   $n_p=0.963\pm0.014$).

2. The SDMC with the same mass $M\simeq10^{-6}M_\odot$ but formed at
   the RD stage from the peak with $\delta_H=0.05$.

3. The limiting case of SDMC: clumps are formed at $t\sim t_{\rm eq}$ from
   $\delta_{\rm eq}\sim1$ perturbations. For example, we consider here
   the mass $M\simeq0.1M_\odot$, which correspond to the DM mass-scale
   of the quark-gluon transition and to the mass of microlensing
   objects.

  \subsection{Non-spherical model}
  
  \label{nonsphsubsec}

In general the shapes of the real perturbations are not spherical. 
To explore the importance of this non-sphericity one must go beyond the
spherical approximation. We model the non-spherical evolution of SDMC
by the homogeneous ellipsoid. Though this model doesn't take into
account the internal structure of the real SDMC, it's useful for the
calculation of the non-spherical outer layers evolution, where the
most of mass is concentrated. For the detailed description of the
nonlinear homogeneous ellipsoid model see \cite{EisLoe95}. We consider
here the small departures from the spherical shape, it will allow us
to simplify the full nonlinear problem, and this approximation is
enough for our purposes.

The total potential of the perturbation
\begin{equation}
\phi=\frac{1}{2}\Phi_{\alpha\beta}(t)r^{\alpha}r^{\beta}
\end{equation}
includes the potentials of ellipsoid (excess of the DM), homogeneous
background and tidal forces:
\begin{equation}
\Phi=\Phi_{\rm el}+\Phi_{\rm bg}+\Phi_{\rm sh}, ~~~ \Phi_{\rm bg}=4\pi
G\bar \rho(t)I/3,
\end{equation}
where $I$ is the unit matrix.
We neglect here the tidal term $\Phi_{sh}$ and a possible
ellipsoid's rotation.  Evolution is described by the equations
\begin{equation}
\frac{d^2 S^{\alpha\beta}}{dt^2}=
-\Phi^{\alpha\gamma}S^{\gamma\beta}.\label{evolq}
\end{equation}  
The matrix of the ellipsoid in the frame of it's main semi-axis can be
written as
\begin{equation}
\begin{array}{cc}
S= \left\|
\begin{array}{ccc}
a& & \\ &b& \\ & &c
\end{array}
\right\|=Ir+\sigma, & \Phi_{\rm el}=2\pi G\rho_e \left\|
\begin{array}{ccc}
A_1& & \\ &A_2& \\ & &A_3
\end{array}
\right\|
\end{array},\label{spe}
\end{equation}
where the density of ellipsoid $\rho_e\equiv M_e/V$ is
given by
\begin{equation}
\rho_e=\rho_m\left(\frac{1+\delta_i}{b^3}-1\right).
\end{equation}
The coefficients of the potential are \cite{EisLoe95}
\begin{equation}
A_1=abc\int\limits_0^{\infty} \frac{d\lambda}
{(a^2+\lambda)[(a^2+\lambda)(b^2+\lambda)(c^2+\lambda)]^{1/2}},
\end{equation}
and $A_2$, $A_3$ can be written in  the similar way. We denote
$\Delta=Tr(\sigma)$ and expand the potential up to the first power in
the variables $\sigma\ll 1$ as
\begin{equation}
\Phi_{\rm el}=2\pi
G\rho_e\left\{\frac{2}{3}\left(1+\frac{2}{5}\frac{\Delta}{r}\right)I-\frac{4}{5}\frac{\sigma}{r}\right\}.
\label{bigphieq}
\end{equation}
At the initial moment $t_i$ we choose the radius $r$ so that
$\Delta(t_i)=0$, then it follows from (\ref{evolq}) that $\Delta(t)=0$
at any time $t$ (similar truck was used e.~g. in \cite{ZabNasPol87}).

In the zero order $\sigma=0$ we have the same Eq.~(\ref{bigeq}). In
the next order we obtain the equation for the $\sigma$
\begin{equation}
\frac{d^2\sigma}{dt^2}=\frac{4\pi}{15} G\rho_e\sigma-\frac{4\pi}{3}
G(2\rho_r+\rho_m)\sigma
\end{equation}
With the parametrization $\sigma=a(y)s(y)\xi$, the equation for the
new function $s(y)$ can be written as
\begin{equation}
y(y+1)s''+\left(1+\frac{3}{2}y\right)s'-\frac{1}{10}
\left(\frac{1}{b^3}-1\right)s=0.\label{bigeqsigma}
\end{equation}

Let us consider the initial conditions for the homogeneous ellipsoid
at the RD-stage in the conformal Newtonian frame. At the scales $r\gg
ct$ one has $\delta_r=-2\Phi=const$,
$\delta_i=(3/4)\delta_{r,i}=-(3/2)\Phi_i$.  The solution for the
relativistic potential is \cite{GorRub10}
\begin{equation}
\Phi(\eta,\vec k)=\Phi_i(\vec
k)\frac{3\pi^{1/2}}{2^{1/2}(u_sk\eta)^{3/2}}J_{3/2}(u_sk\eta),
\end{equation}
where $u_s=1/\sqrt{3}$. The peculiar velocities $v_j$, which define
the initial velocity of the ellipsoid's surface, are expressed as
$v_j=\partial v/\partial x_j$, where the velocity potential
\cite{GorRub10}
\begin{equation}
v(\vec k)=-\frac{1}{\eta}\int\limits_0^\eta d\eta'\eta'\Phi(\eta',\vec
k)=-9\Phi_i(\vec k)\frac{1}{\eta k^2}.
\end{equation}

The ellipsoidal top-hat perturbation has the form $\delta_i(\vec
x)=\delta_i=const$ if $(x/a)^2+(y/b)^2+(z/c)^2\leq1$, and
$\delta_i(\vec x)=0$ otherwise. The Fourier transform of the
ellipsoidal top-hat is \cite{KoaSarOza07}
\begin{equation}
\delta_i(\vec k)=\delta_i(2\pi)^3abc\left(\frac{\sin(\tilde k)-\tilde
k\cos(\tilde k)}{2\pi^2\tilde k}\right),
\end{equation}
where $\tilde k=[(ak_x)^2+(bk_y)^2+(ck_z)^2]^{1/2}$. Let us denote
$\tilde {\vec x}=(x/a,y/b,z/c)$, then
\begin{equation}
v(\tilde {\vec x})=\frac{1}{abc}\int\frac{d^3\tilde
k}{(2\pi)^3}\frac{-9\Phi_i(\vec k)e^{-i\tilde {\vec x}\tilde {\vec
k}}}{\eta\left[(\tilde k_x/a)^2+(\tilde k_y/b)^2+(\tilde
k_z/c)^2\right]}.
\end{equation}
In the approximation of small nonspericity one can write
\begin{equation}
v=v_0+\left.\frac{\partial v}{\partial a}\right|_0\Delta
a+\left.\frac{\partial v}{\partial b}\right|_0\Delta
b+\left.\frac{\partial v}{\partial c}\right|_0\Delta c+\cdots,
\label{vser}
\end{equation}
where zero corresponds to the spherical zero-order case with
$a=b=c$. The only nonzero derivative  $v_j=\partial v/\partial x_j$ is
over the correspondent axis $x_j$ for the linear terms in
(\ref{vser}). After some algebra we have the initial conditions in the
form
\begin{equation}
\left.s\right|_{y_i}=s_i \mbox{~~~~and~~~~}
\left.s'\right|_{y_i}=\frac{3\delta_Hb_i^3s_i} {10y_i\phi}
\end{equation}
for each component of $s$.

\begin{figure}[t]
\begin{center}
\includegraphics[angle=0,width=0.45\textwidth]{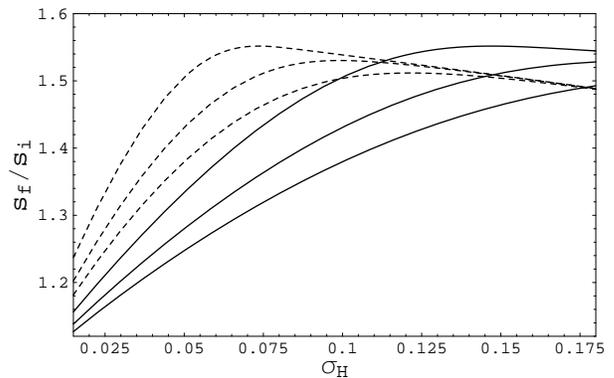}
\end{center}
\caption{Asphericity growth $s_f/s_i$ in dependence on the
r.m.s. perturbation value $\sigma_H$ at the horizon crossing for
different clumps' masses $M=10^{-6}$, $10^{-1}$, $10^{2}M_\odot$ (from
up to down). The solid and the dashed curves corresponds to the
$\nu=1$ and $\nu=2$ peak hights, respectively, where $\nu\equiv\delta_H/\sigma_H$.} \label{gran}
\end{figure}

We solve Eg.~(\ref{bigeqsigma}) numerically simultaneously with
(\ref{bigeq}) for the peak highs $\nu=1$ and $\nu=2$, where $\nu$ is defined as $\nu\equiv\delta_H/\sigma_H$ and $\sigma_H\equiv\langle\delta_H^2\rangle^{1/2}$. The results are presented at the Fig.~\ref{gran}. 
We calculated the grow of the asphericity till the detachment of
the object from the Hubble flow.   Note that in the case of
isocurvature perturbation the similar formalism (with initial
$s'(t_i)=0$) shows that the asphericity growth is small, $s$ changes
by less then 10\%, therefore for the isocurvature perturbation the
asphericity constraints are not strong. Now we return to the curvature
perturbations. The boundary of the allowed     asphericity is
roughly $\Delta s_f/b_f<1$. From this condition the following
criterion for the SDMC formation follows $\Delta
s_i/b_i<(b_f/b_i)(\Delta s_i/\Delta s_f)$. The perturbations with
larger $\Delta s_i/b_i$ will not form the clumps at the RD-stage. 

\begin{figure}[t]
\begin{center}
\includegraphics[angle=0,width=0.45\textwidth]{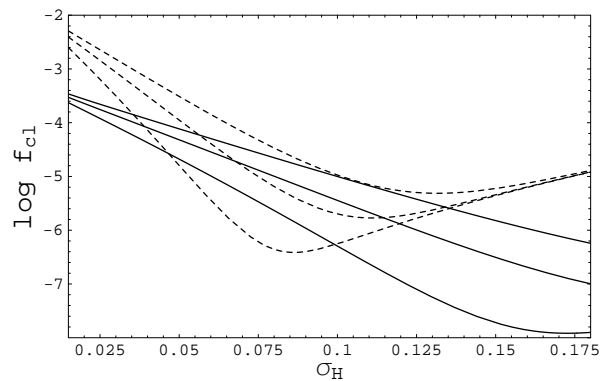}
\end{center}
\caption{Fraction of clumps $f_{\rm cl}$ which formed by surviving the
growth of anisotropy for clumps masses $M=10^{2}$, $10^{-1}$,
$10^{-6}M_\odot$ (from up to down). The solid and the dashed curves
corresponds to the $\nu=1$ and $\nu=2$ peak highs, respectively. }
\label{fclgr}
\end{figure}

  Now let us consider the distribution of the initial asphericity of perturbations. The
distribution of Gaussian perturbation over their shapes can be
calculated with the help of results of \cite{Dor70}, \cite{Baretal86}.   With the help of (\ref{bigphieq}) the ellipticity of the potential of
the ellipsoidal distribution can be expressed as
\begin{equation}
e=\frac{\lambda_1-\lambda_2}{2\sum\lambda_i}\simeq\frac{1}{5}\frac{\Delta
s_i}{b_i},
\label{elambda}
\end{equation}
where $\lambda_i$ are the eigenvalues of the gravitational-shear
tensor. The 2nd variable prolateness is expressed as
\begin{equation}
p = \frac{\lambda_1+\lambda_3-2\lambda_2}{\sum\lambda_i} .
\end{equation}
The distribution over eigenvalues $\lambda_1\ge \lambda_2\ge
\lambda_3$ was found in \cite{Dor70} in the form
\begin{eqnarray}
p(\lambda_1,\lambda_2,\lambda_3) &=& {15^3\over
8\pi\sqrt{5}\,\sigma^6}\  \exp\left(-{3 I_1^2\over \sigma^2} + {15
I_2\over 2\sigma^2}\right) \times \nonumber \\
&\times&(\lambda_1-\lambda_2)(\lambda_2-\lambda_3)(\lambda_1-\lambda_3),
\label{lambdas}
\end{eqnarray}
where $\sigma$ is the r.s.m. perturbation,  $I_1=\lambda_1 + \lambda_2
+ \lambda_3$, and  $I_2=\lambda_1\lambda_2 + \lambda_2\lambda_3
+\lambda_1\lambda_3$.  This distribution can be expressed through $e$
and $p$ variables as \cite{SheMoTor01}
\begin{equation}
g(e,p|\nu) = {1125\over \sqrt{10\pi}}\,e\,(e^2-p^2)\, \nu^5 {\rm
e}^{-{5\over 2}{\nu^2}(3 e^2 + p^2)},
\label{distrep}
\end{equation}  
where $\nu\equiv\delta/\sigma$. We use this distribution, integrated over
$p$ in the range $-e<p<e$, for the calculations of the fraction of the
formed SDMC. The results are shown at Fig.~\ref{fclgr}. Therefore the
effect of asphericity can diminish the number of the formed SDMC by
several orders.

\begin{figure}[t]
\begin{center}
\includegraphics[angle=0,width=0.45\textwidth]{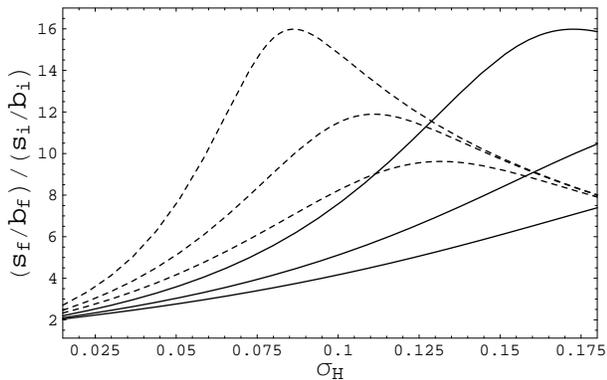}
\end{center}
\caption{  The same as at Fig.~\ref{gran} but for the relative asphericity growth $(s_f/b_f)/(s_i/b_i)$.  } \label{granrel}
\end{figure}

  The smallness of $f_{\rm cl}$ can also be explained by the following way. The absolute values of 
$s$ grows only moderately, and the fraction $s_f/s_i$ is not large. But the relative asphericity grows significantly because of the large value of $b_i/b_f$. The ratio of $s_f/b_f$ and $s_i/b_i$ is shown at Fig.~\ref{granrel}. Therefore the initial asphericity $s_i/b_i$ should be very small for the final asphericity to be $s_f/b_f<1$. The probability distribution depends namely on the relative asphericity $s/b$. This is the reason why the initial distribution of asphericity (\ref{distrep}) gives the huge suppression of the formed clumps fraction.


\section{Secondary accretion and the formation of ultracompact minihaloes}

  \subsection{Smooth initial perturbation v.s. compact seed}
\label{genclumpssub}

There are two distinct but complementary models of the clumps
formation: evolution of the initially smooth perturbation, and the
secondary accretion onto the previously formed seed. Let us
consider the dust-like stage of the universe. The global parameters of
the formed clump can be found from the simple top-hat model
\cite{svvbm}. From the linear perturbation growth theory taking into
account the gradual transition from the radiation to matter-dominated
stage the expression follows \cite{GorRub10}
\begin{equation}
\delta(k,z)\simeq\frac{27}{2}\Phi_i(k)\frac{1+z_{\rm
eq}}{1+z}\ln(0.2k\eta_{\rm eq}),
\label{deltalog}
\end{equation}
where $\Phi_i$ is the initial gravitational potential of the
perturbation well outside the horizon.   The formation time $t_c$ of
the clump can be found from the relation  $\delta(t_c)=\delta_c$,
where  $\delta_c=3(2\pi)^{2/3}/20\approx1.686$ (see
e.~g. \cite{LacCol93}). The mean density of the clump is
$\bar\rho_{\rm int}=\varkappa \bar\rho(t_c)$, where
$\varkappa=18\pi^2\approx178$, and its virial radius
\begin{equation}
R=\left(\frac{3M}{4\pi\bar\rho_{\rm int}}  \right)^{1/3}.
\label{rad1}
\end{equation}
Further growth of the clump is due to virialization of the new
spherical layers, if the perturbation extends to the larger scales
till the regions, where $\delta=0$ -- this is the boundary of the
perturbation region with total energy $E<0$. The statistics of the
clumps is determined by the power-spectrum of perturbations,
normalized on the observational data. One can use in the
(\ref{deltalog}) the  spectrum of potential ${\mathcal P}_\Phi$ for
which $A_\Phi=(4/9)A_{\mathcal R}$ in (\ref{prspectrum}).

  If these clumps were formed from the isolated density
perturbations, they probably have the Gurevich-Zybin density profiles
$\rho\propto r^{-1.8}$ \cite{GurZyb8895-1}, \cite{GurZyb8895-2}, \cite{GurZyb8895-3}
or near isothermal profile $\rho\propto r ^{-2}$  \cite{Vogetal09}.
The rather different profiles arise in the hierarchical clustering, if the clump was formed by the aggregation of smaller clumps. In this case the clump probably have the NFW or similar density profiles.

Let us discuss the secondary accretion scenario in more
details. In the spirit of secondary accretion models, we assume that at the
RD-stage there is the compact seed mass $M_c$, and the DM is
distributed homogeneously around it. For $t\ll t_{\rm eq}$ and for the
mass $M\gg M_c$, the isocurvature perturbation of DM $\delta_i=M_c/M $
does not evolve. Indeed, according to the Meszarosh solution
$\delta=\delta_i(1+3x/2)$  \cite{svvbm}, where $x=a/a_{\rm eq}$. This
solution can  easily be obtained from equation (\ref{bigeq}) in the
linear approximation.

The secondary accretion begins at $t\sim t_{\rm eq}$. If we assume
that the Hubble flow was not perturbed (such situation takes place for
entropy perturbations), then one have to replace the
$5\delta/3\to\delta$ in the top-hat model \cite{Ber85}, and the
threshold for the object formation in this case is
$\delta(t)=\tilde\delta_c=(3\pi/2)^{2/3}\approx2.81$. Using the
Meszarosh solution one finds the mass $M$ of the virialized object in
dependence of the redshift
\begin{eqnarray}
M(z)&=&\frac{3}{2}\left(\frac{2}{3\pi}\right)^{2/3}
\frac{1+z_{\rm eq}}{1+z}M_c
\label{rcol2}
\\
&\approx&1.7\times10^3\left(\frac{M_c}{10^2M_{\odot}}\right)\left(\frac{1+z}{100}\right)^{-1}M_\odot, \nonumber
\end{eqnarray}
where $(3/2)(2/(3\pi))^{2/3}\approx0.53$. This numerical
coefficient is about $1/2$ of the coefficient obtained in the work
\cite{RicGou09}. The virial radius of this clump is
\begin{eqnarray}
r_c&=&\frac{1}{3}\left(\frac{3}{4\pi} 
\right)^{1/3}
\frac{M^{4/3}}{\rho_{\rm eq}^{1/3}M_c}
\label{rcol}
\\
&\simeq&
3.2\left(\frac{M_c}{10^2M_{\odot}}\right)^{1/3}
\left(\frac{100}{1 + z}\right)^{4/3}\mbox{~pc}.\nonumber
\end{eqnarray}
Using the connection  (\ref{rcol}) between the halo mass and its
radius, one finds the density profile of the ``induced halo'' at the
distances where $M>M_c$:
\begin{eqnarray}
\rho(r)&=&\frac{1}{4\pi r_c^2}\left.\frac{dM(r_c)}{dr_c}\right|_{r_c=r}
\label{rhoih}
\\
&\simeq &3\times10^{-21}\left(\frac{r}{1\mbox{~pc}}\right)^{-9/4}\left(\frac{M_c}{10^2M_\odot}\right)^{3/4}\mbox{g~cm$^{-3}$}.
\nonumber
\end{eqnarray}
In the case of non-compact central object, such
as extended clusters of
primordial black holes, the density profile does not match exactly the
profile $\rho\propto r^{-9/4}$.

  \subsection{Superdense clumps as the seeds}

The UCMHs can be formed around the existing SDMC.  The structure
of a UCMH around a SDMC, originated from the entropy perturbations,
was considered in \cite{KolTka94}. There is no principle difference
with the SDMC originated from the peaks in the spectrum of the
adiabatic perturbations. In both cases the rapid growth of UCMH
begins only at $t\sim t_{\rm eq}$.

In some sense a SDMC plays the role of the core radius of the
UCMH. It's important that the secondary accretion density profile
$\rho\propto r^{-9/4}$   definitely continues inward 
down to the radius of the SDMC. At the 
smaller radius the density profile of the UCMH is the
SDMC's profile which may be the Gurevich-Zybin profile $\rho\propto r^{-1.8}$ 
with some core, although steeper profiles are also 
possible. In any case some mechanism may stop the density growth at the center of the SDMC (see Section~\ref{coresec}).

  \subsection{Final mass of the ultracompact minihalos}
  
Total mass of induced halo increases with time as more and more remote
areas around the seed are separated from the cosmological expansion
and virialized. The growth of the induced halo is terminated in an era
of the nonlinear stage  development of the normal inflationary
perturbations of DM with mass which is of the order of the growing
mass $M(t)$ of the induced halo at that epoch. The growth laws
$\delta\propto t^{2/3}$ of the normal perturbations and the
perturbations induced by the seed mass, are the same in the era of the
matter dominance.

\begin{figure}[t]
\begin{center}
\includegraphics[angle=0,width=0.45\textwidth]{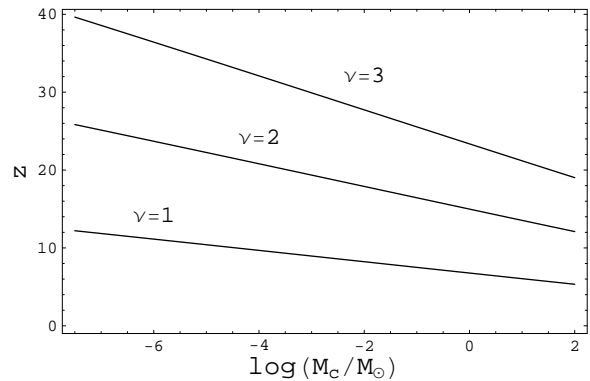}
\end{center}
\caption{  The red-shift $z$ of the growth termination according to Eq.~\ref{Deltaeq} for $\nu=1$, $2$ and $3$ fluctuations highs.  } \label{term1}
\end{figure}

\begin{figure}[t]
\begin{center}
\includegraphics[angle=0,width=0.45\textwidth]{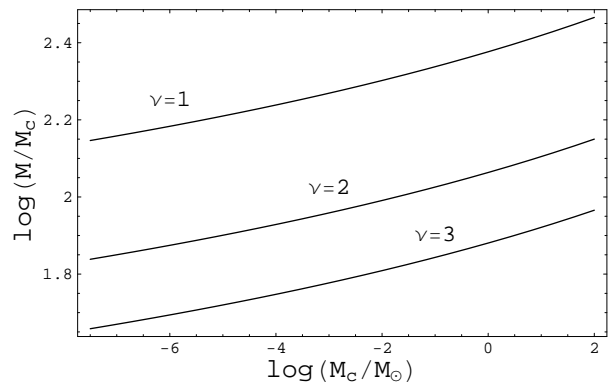}
\end{center}
\caption{  The final mass of the secondary accreted halo $M$ in relation to the seed mass $M_c$ for $\nu=1$, $2$ and $3$ fluctuations highs.  } \label{term2}
\end{figure}

The particular UCMH is surrounded by several perturbations of the same
mass-scale. At about half of them the perturbation is
positive. Therefore, the condition of the growth ceasing from the side of
positive perturbation is
\begin{equation} 
\label{Deltaeq}
\nu\sigma_{\rm eq}(M)=\frac{9}{10}\frac{M_c}{M},
\end{equation}
where $\nu$ is the value of the density perturbation in terms of rms
fluctuation $\sigma_{\rm eq}(M)$. The right-hand side of equation
(\ref{Deltaeq}) denotes the amplitude of the fluctuations caused by
the mass $M_c$ with the factors discussed in the
Section~\ref{genclumpssub}. The left side of the equation represents
the usual Gaussian fluctuations at the time  $t_{\rm eq}$. Due to the
large number of the surrounding perturbations the perturbations with
the positive $\nu\sim1$ will stop the UCMH growth from several
directions and will destroy the regular secondary accretion. In
addition, the negative perturbations $\nu<0$ will also dump the UCMH
growth due to the lower density of DM inside them. Numerical solution
of (\ref{Deltaeq}) gives the final average mass of the induced halo
$M\sim(10^{1.5}-10^{2.5})M_c$ in the range $M_c\sim10^{-8}-10^2M_\odot$.  The mechanism of
the induced halo's growth stop, similar in some aspects to the one
discussed in this section, was used in the works \cite{DokEro01},
\cite{DokEro03}, \cite{MacOstRic07} for the calculation of the induced
halo properties.    The red-shift of the growth termination and the fraction $M/M_c$, found from the relations (\ref{Deltaeq}), $\delta(t_c)=\delta_c$ and known function $z(t)$, are shown at the Fig.~\ref{term1}, Fig.~\ref{term2}, respectively. The calculated redshift is roughly consistent with $z\sim 10-30$ predictions.  


\section{Maximum density in the centers of the clumps}
\label{coresec}

The density inside the clump grows towards the center. If the density stops
growing at some radius $r=R_c$, then the region $r<R_c$ is referred to
as ``core''. The evidences for the relative core radius
$R_c/R\sim10^{-2}$, where $R$ is the virial radius, are seen in the
numerical simulations \cite{DieMooSta05}. However, in another simulation
\cite{IshMakEbi10} the core is not seen
down to $R_c/R\sim10^{-3}$; in this
simulation the power-law growth continues down to the distance,
where the numerical resolution fails. Let us discuss several
possible restrictions on the maximum density in the clumps centres
(core radius).

  \subsection{Liouville's theorem restriction for the core radius}
  \label{Lauwilethermsub}

The Liouville's theorem is known in the two equivalent formulations:
the phase volume $\int dqdp$ is conserved or the distribution function
in the phase space $f(p,q)=const$. Therefore the probability density
at the core (mean value of $f(p,q)$ over core volume) cannot exceed
initial value due to Liouville's theorem. The only necessary condition
for the theorem validity is the Hamiltonian character of the
system. Unfortunately,  $R_c$ can only be restricted, but not
determined with the help of the Liouville's theorem applied only to
the initial and final stages of clump evolution. This is because the
entropy production in the intermediate processes
\cite{DorLukMik12}. By other words, the phase volume becomes dispersed
during the evolution (see Fig.~8.3 in \cite{GorRub10}).

The distribution function can be estimated simply as $f_c=\rho_c/v^3$,
where $\rho_c$ is the core density and $v$ is the characteristic
velocity in the core. For the near isothermal density profile
$\rho(r)\propto r^{-2}$ the $v$ is of the order of the virial velocity
of the whole clump. We restrict the $R_c$ from the Liouville's theorem
applied to the initial and final stages.

There are two sources of initial entropy or initial $\sigma$: thermal
velocities of DM particles at decoupling and peculiar velocities in
the case of adiabatic density perturbations.

The thermal part can be attributed to the distribution function at the
time of kinetic decoupling $t_d$ \cite{GorRub10} (Section 8.3.2). The
neutralino is nonrelativistic nondegenerate at $t_d$, therefore the
good approximation for the distribution function at this moment is the
Maxwell's distribution:
\begin{equation}
f_p(p)d^3rd^3p=\frac{\rho_m}{m(2\pi
mkT)^{3/2}}e^{-\frac{p^2}{2mkT}}d^3rd^3p,
\label{maxdistr}
\end{equation}
where $\rho_m$ is the density of DM, which is expressed through the
temperature at any time by the using of the entropy conservation
condition $g_*T^3a^3=const$, where $g_*$ is the effective number of
degrees of freedom at the temperature $T$, and $m$ is the mass of DM
particle.  The distribution function inside the core is less then the
initial distribution function, which has the maximum value at
$p=0$. Therefore we use the inequality $f_c<f_p(p=0)$.  For the
isothermal density profile in the clump this condition gives the
restriction on the  relative core radius
\begin{equation}
\frac{R_c}{R}>\frac{2\pi^{1/2}\bar\rho^{1/4}T_d^{3/4}}{3^{1/4}G^{3/4}M^{1/2}m^{3/4}\rho_m^{1/2}(t_d)}.
\label{rcliwtherm}
\end{equation}
The numerical examples for the case of $m=100$~GeV neutralinos with
the temperature of the kinetic decoupling $T_d\simeq25$~MeV are shown
in the Table~\ref{tab}.

Now we turn to the peculiar velocities, which are generated due to the
gravitational instability and can play the role analogous to the
thermal velocities in the Liouville theorem restriction.   
Normally, the cut-off of the spectrum is not sharp. Even for the usual exponential cut off, e.g., the $2k$ modes are present with sufficiently large amplitudes, where $k$ is the cut-off scale. The $2k$ modes give the peculiar velocities which are distributed near isotropicaly, compared with the approximately radial $k$ modes. Therefore, the modes $k'>k$ can be considered as approximately randomly directed in analogy with the thermal velocities in the Liouville theorem.

The peculiar velocity at some mass scale in the case of the flat metrics can be
expressed as \cite{svvbm}
\begin{equation}
\vec v=\frac{Ha}{4\pi}\nabla_x\int\frac{d^3x'\delta(\vec x')}{|\vec
x'-\vec x|}.
\label{vnabla}
\end{equation}
The growing mode comprises only $2/5$ fraction of (\ref{vnabla}). Just
after the $t_{\rm eq}$ the falling mode ceased, the peculiar
velocities developed and grow later as $\propto t^{1/3}$. We take the
initial stage at $t=t_{\rm eq}$. The calculations are analogous to the
thermal case with the replacement of the thermal velocity by the
peculiar one. We obtain the minimum relative core radius in the form
\begin{equation}
\frac{R_c}{R}=0.01\delta_{\rm eq}^{9/2}.
\label{rcpec}
\end{equation} 
The numerical results of the calculations (\ref{rcliwtherm}) and
(\ref{rcpec}) are shown in the Table~\ref{tab}. At the RD-stage the
peculiar velocities are estimated according to the formalism presented
in the Section~\ref{formsec}.

  \subsection{Annihilation criterion}
  
In the work \cite{Ulletal02} the maximum density in the center of the
clump was estimated from the annihilation rate and the elapsed time:
\begin{equation}
\rho(r_{\rm min})\simeq \frac{m}{\langle\sigma v\rangle(t_0-t_f)}
\label{annwr}
\end{equation}
where $t_0$ is the current moment of time and $t_f$ is the formation
moment of the clump. According to this estimate the core region
becomes larger due to the loss of the particles at the orbits, that
goes through the center of the clump. For the isothermal
$\rho(r)\propto r^{-2}$ density profile, the corresponding relative
core radius
\begin{equation}
\frac{R_c}{R}\simeq\left(\frac{\langle\sigma v\rangle
t_0\bar\rho}{3m}\right)^{1/2},
\label{rcann1}
\end{equation}
is presented in the Table~\ref{tab} for the thermal production value
$\langle\sigma v\rangle\simeq3\times10^{-26}$~cm$^3$~s$^{-1}$ and
$m=100$~GeV. This approach assumes that the orbits passing through
the center are not fulfilled after the particles 
annihilation.

The opposite case with the compensation of the particles loss was
considered in \cite{BerGurZyb92} and \cite{BerBotMig96}. In
\cite{BerGurZyb92} it is found that  the core radius is defined by 
the annihilation
at the stage of the halo formation. The minimum radius was found from
the condition that the annihilation time is of the order of the Jeans
time, because this time defines the characteristic time of the density
profile formation. This corresponds to the epoch of the halo
formation. In the article \cite{BerBotMig96} the core radius of the
already formed clump was found in the assumption of the steady
hydrodynamic bulk flow of the DM to the clump center. It was putted
that the free fall time is equal to the annihilation time, and the
center of the clumps is always fulfilled by the DM particles from the
flow.    

In the real clump there are mechanisms of the regeneration of the
orbits with the small angular momentums, which goes through the center
of the clump. These conditions result in the larger central
density in comparison with (\ref{annwr}). These orbits exist in
some degree due to the clump's dynamical restructuring. Even the tidal
forces from nearby stars and gravitational shocks at the galactic disk
crossings lead to the changes of the clumps dynamical structure and
appearance of the orbits with small angular momentum. The exact
problem treatment requires the solution of the self-consistent kinetic
equations for the clump structure with the influence of the external
tidal forces.

  \subsection{Possible mechanisms of the core formation} 

In the above subsections the lower limits on the relative core radius
were presented.   It's clear from the Table~\ref{tab}, that the
Liouville restriction with the peculiar velocities can be the dominant
effect in the formation of the core in the very dense clumps, formed
just after the matter-radiation equality.

Can other effects produce the core with the greater sizes? Dwarf and
LSB galaxies tell us that some unknown effect is responsible for their
large cores, which was not reproduced yet in the N-body
simulations. First of all, the phase-space density is diluted by the
entropy generation during the nonlinear hierarchical clustering
\cite{DorLukMik12}. In the work \cite{DorLukMik12} this effect was
explored for the galaxies-sized halos, and \cite{DorLukMik12}
concluded that at the small scales the entropy generation is less
effective, but the quantitative value is unclear.

Another affect is the tidal forces, which deflect particles, defocus
them from the center of the clump at the stage of it's formation
\cite{BerDokEro03}. May be, similar effect define the break of the
density profile in the clump \cite{SikTkaWan96}. Interesting analytic
estimation of the core radius $R_c/R\simeq\delta_{eq}^3$ was obtained
in \cite{GurZyb8895-1}, \cite{GurZyb8895-2}, \cite{GurZyb8895-3} from the energy criterion and the falling mode
influence. But the real mechanism of the core formation is still
unclear.

\begin{widetext}
\begin{center}
\begin{table}[tbp]
\centering
\begin{tabular}{|c|c|c|c|c|c|c|}
\hline
 $M/M_\odot$  & $\bar\rho$, g~cm$^{-3}$   &  $\delta$ &
$R_c/R$, Liouv. therm.  & $R_c/R$, Liouv. pec.   & $R_c/R$,
annih.  \\
\hline 
$10^{-6}$ & $3\times10^{-23}$   & $\delta_{\rm eq}=0.009$
&  $4\times10^{-3}$  &  $6\times10^{-12}$  & $2.6\times10^{-5}$  \\
\hline
$10^{-6}$  & $4.2\times10^{-16}$   &  $\delta_H=0.05$ & 0.24  &
0.1 & 0.1  \\
\hline 
$0.1$  &  $2.5\times10^{-17}$  & $\delta_{\rm eq}\simeq 1$
&  $4\times10^{-4}$  & 0.01   & $2.5\times10^{-2}$  \\
\hline
\end{tabular}
\caption{\label{tab} The parameters of the clumps in the three examples
and the relative core radius due to different effects.}
\end{table}
\end{center}
\end{widetext}


\section{Annihilation of DM in the ultracompact minihalos}
\label{annsec}

Let us estimate the annihilation signal from the UCMHs seeded by the
SDMCs. The annihilation rate in the single UCMH contains contributions
from the three components, the SDMC $r<R$, the intermediate region
$R\leq r\leq R_e$ and the outer halo $R_e\leq r\leq R_h$ formed by the
secondary accretion at the dust-like stage,
\begin{equation}
\dot N_{\rm
cl}=4\pi\left[\int\limits_{0}^{R}+\int\limits_{R}^{R_e}+\int\limits_{R_e}^{R_h}
\right] r^2dr\rho_{\rm int}^2(r) m^{-2}\langle\sigma_{\rm ann}
v\rangle,
\label{sepfour}
\end{equation}
where $\sigma_{\rm ann}$ is the annihilation cross-section. The first
integral in the square-brackets is equal to
\begin{equation}
\dot N_{\rm cl}^{(1)}=\frac{3}{4\pi} \frac{\langle\sigma_{\rm ann}
v\rangle}{m^2} \frac{M^2}{R^3}\,S, \label{separ}
\end{equation}
where the function $S$ depends on the distribution of DM, and $S=1$ in
the simplest case of the uniform density $\rho_{\rm int}(r)=const$.
For the power-law density distribution $\rho\propto r^{-\beta}$ with
relative core radius $x_c=R_c/R$ we have
\begin{equation}
S(x_c,\beta)\!=\! \frac{(3-\beta)^2}{3(2\beta-3)}
\left(\frac{2\beta}{3}x_c^{3-2\beta}-\!1\!\!\right)\!\!
\left(\!1\!-\!\frac{\beta}{3}x_c^{3-\beta}\right)^{-2}.
\label{sbig}
\end{equation}
In particular $S\simeq 4/(9x_c)$ for the profile  $\rho\propto r^{-2}$
with $x_c\ll 1$.  If $\beta\geq1.5$ the contribution (\ref{separ}) is
dominant in the (\ref{sepfour}), so one can neglect the two last
integrals in (\ref{sepfour}). Note, however that in the case of UCMH
with PBH seed the second integral would be dominant.

We parametrize the annihilation signal in the $\psi$ direction (with
respect to Galactic center) over solid angle $\Delta\Omega$ by the
usual way \cite{BerEdsGonUll99}:
\begin{equation}
J_{\gamma}(E,\psi,\Delta\Omega)=9.4\times10^{-11}\frac{dF}{dE}\langle
J(\psi)\rangle_{\Delta\Omega}, \label{igam}
\end{equation}
where
\begin{equation}
\frac{dF}{dE}=\left(\frac{100\mbox{~GeV}}{m}\right)^2\sum
\limits_{F}\frac{\langle\sigma_Fv\rangle}{10^{-26}\mbox{~sm$^3$s$^{-1}$}}
\frac{dN_{\gamma}^F}{dE}, \label{dsde}
\end{equation}
and the astrophysical factor
\begin{equation}
\langle
J(\psi)\rangle_{\Delta\Omega}=\frac{1}{8.5\mbox{~kpc}}\frac{1}{\Delta\Omega}
\int d\Omega'\int
dL\left(\frac{\rho(r)}{0.3\mbox{~GeV~sm$^{-3}$}}\right)^2, \label{jpsi}
\end{equation}
where the integration goes along the line-of-sight. For the diffuse
signal from the clumped DM
\begin{eqnarray}
&~&\langle J(\psi)\rangle_{\Delta\Omega}= \label{jpsi-pi}
\\
&~&f_{\rm cl}S\left(\frac{\bar\rho_{\rm int}}{0.3\mbox{~GeV~sm$^{-3}$}}\right)
\int\limits_{l.o.s.}\frac{dL}{8.5\mbox{~kpc}}\left(\frac{\rho_H(r)}
{0.3\mbox{~GeV~sm$^{-3}$}}\right),\nonumber
\end{eqnarray}
where for the Galactic halo density profile $\rho_H(r)$ we use the NFW
profile. For the assumed annihilational canal the gamma-rays are
generated due to the pion productions and decays
$\pi^0\to2\gamma$. Denote by $\eta_{\pi^0}\sim10$ the number of
photons per one decay.

\begin{figure}[t]
\begin{center}
\includegraphics[angle=0,width=0.45\textwidth]{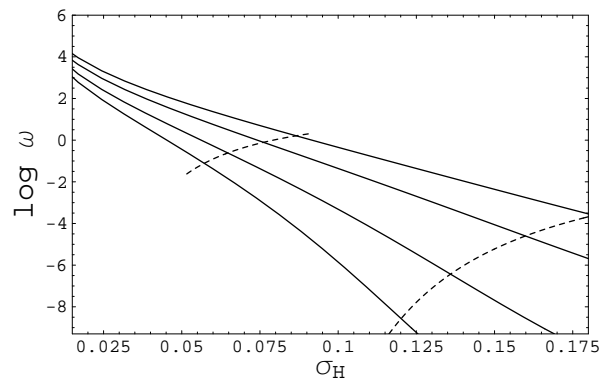}
\end{center}
\caption{Solid lines show the upper limits on the parameter $\omega$
in the Eq.~(\ref{omdef}) for the clump masses $M=10^{2}$, $10^{-1}$,
$10^{-6}M_\odot$ and $10^{-11}M_\odot$ (from up to down).  The PBH
restrictions with the thresholds $\delta_{\rm th}=1/3$ and
$\delta_{\rm th}=0.7$ are shown by the left and right dashed curves,
respectively.} \label{grom}
\end{figure}

The flux of gamma-rays from the annihilation may exceed the Fermi-LAT
limits for some regions of the free parameters space. Here we consider
the annihilation of the ordinary $m\sim100$~GeV neutralino. There are
several uncertain parameters in the considered problem: core radius in
the seed SDMC, mass $m$ and annihilation cross-section. We estimated the
survived SHDM fraction $f_{\rm cl}$ in the Section~\ref{nonsphsubsec},
but this value is very sensitive to the boundary of the allowed
anisotropy $\Delta s_f/b_f<1$. Possibly, the boundary will be find in
the future N-body simulation of the SDMCs formation. At the current
level we still consider $f_{\rm cl}$ as the free parameter. Let us
denote by the $\omega$ the following combination of parameters
\begin{equation}
\omega=\left(\frac{\eta_{\pi^0}}{10}\right)\left(\frac{m}{100\mbox{~GeV}}\right)\left(\frac{\langle\sigma
v\rangle}{10^{-26}\mbox{~cm$^3$s$^{-1}$}}\right)\left(\frac{f_{\rm
cl}}{10^{-5}}\right)S.\label{omdef}
\end{equation}
We compare the calculated annihilation signal in the Galactic
anti-center direction with the Fermi-LAT diffuse extragalactic
gamma-ray background $J_{\rm
obs}(E>m_{\pi^0}/2)=1.8\times10^{-5}$~cm$^{-2}$~s$^{-1}$~sr$^{-1}$
\cite{Fermi3603}. It gives the conservative limit. The calculated
upper limits on the parameter $\omega$ are shown at
Fig.~\ref{grom}. As one can see, the annihilation of ordinary
neutralino in the SDMC is possible if one takes the effect of
asphericity  from Fig.~\ref{fclgr} into account. Without this effect
one would has $f_{\rm cl}\sim1$, and the annihilation would exceed the
Fermi-LAT limits even in the case of minimum possible annihilation
cross-section which was found in \cite{BerBotMig96}.

\section{Conclusion}
\label{conclsec}

In this paper we discussed several principle aspects of the superdense
dark matter clumps (SDMC) and ultracompact minihalos (UCMH). In some 
particular cases we have calculated their properties. The range
of the free parameters (first of all, describing unknown spectrum of
perturbations at the small scales) are very broad, and it is hardly possible
now to obtain the strict predictions. As a rule, only some restrictions 
are obtained from the annihilation or microlensings \cite{LiEriLaw12}, \cite{Zacetal12}  
limits. The aim of this paper is to find the different types
of DM clumps and their properties to avoid the mishmash, 
which are present occasionally in the literature about UCMH.

The DM clumps can originate at the RD-stage of the Universe evolution, 
as it was demonstrated first by \cite{KolTka94} in the case of the 
entropy perturbations (see also some developments in \cite{KhlSakSok99} and \cite{Khl07}). High peaks (if they exist) atop the spectrum of the
adiabatic perturbation can produce SDMC too \cite{DokEro02}. During 
the SDMC evolution the asphericity of the
perturbation can increase making the object highly non-spherical. It
can prevent the SDMC  formations. But the initial asphericity has a
statistical distribution. In this paper we model the non-spherical objects as 
homogeneous ellipsoids in the approximation of a small asphericity.
The calculated fraction of the formed clumps
is in the range $f_{\rm cl}\sim 10^{-8}-10^{-2}$ depending on the value of 
perturbation at the horizon crossing $\delta_H$ and on mass-scale $M$, 
see Fig.~\ref{fclgr}.

Virialisation at the RD stage proceeds similar to the virialization at
the dust-stage if the clump detached from the cosmological expansion
and its DM density exceeds locally the mean radiation density. The
difference exists only for the very elongated fluctuations. 
Such fluctuations can be ripped apart and disintegrated without the SDMCs formation. 

At the dust stage the secondary accretion can form UCMHs around the
seed PBHs or SDMCs. The induced halos around these seeds grow until
they begin to feel the surrounding density fluctuations. According to our calculations, the final 
mass of the UCMH exceeds the seed mass approximately by two orders.  

The maximum density in the center of a clump may be restricted by
some mechanisms. The Liouville theorem and annihilation criterion give
only the upper limits on the central density, and these limits are
particle-model dependent. In this work we calculated the Liouville theorem's limits
for the thermal and peculiar velocities as the initial data. We have shown that the
Liouville restriction can be the dominant effect for the formation of the core in those clumps, which originated
just after the matter-radiation equality. In the case of SDMCs the annihilation criterion can also restrict the core radius. It's possible that there is some pure
gravitational mechanism like tidal forces or entropy generation, which
limits the central density and produces the central core. But  such a
mechanism is not reliably established.

The observational manifestation of the clumps is due to the 
annihilation of the DM particles inside them, see e.~g. \cite{BelKirKhl12}. 
The particular calculations of the gamma-ray signal require the choice 
of some particular model and parameters for the clumps and DM particles.
As we have shown, the initial asphericity diminishes the number of the 
formed SDMCs and lowers the anticipated annihilation signals from them. Even 
annihilation of ordinary $\sim 100$~GeV neutralinos in the SDMCs without 
exceeding of the Fermi-LAT observational limits is possible if one 
takes the effect of asphericity into account. By using the Fermi-LAT
flux as the upper limit, we obtained the new restrictions on the 
combination of parameters for SDMCs and DM particles, see Fig.~\ref{grom}.

\acknowledgments

Authors thanks A.G.~Doroshkevich and Referee for useful comments. The study was
supported by the Ministry of education and science of Russia, project
8525; and in part by the grants OFN-17, and NSh-871.2012.2.

\end{document}